Running Head: ANIMAL INNOVATION VERSUS HUMAN INNOVATION

# Can Sol's Explanation for the Evolution of Animal Innovation Account for Human Innovation?

Invited commentary on

'The evolution of innovativeness: exaptation or specialized adaptation?' by Daniel Sol.

Liane Gabora and Apara Ranjan

University of British Columbia

For correspondence:

Liane Gabora

liane.gabora@ubc.ca

Department of Psychology, University of British Columbia

ARTS Building, 1147 Research Road

Kelowna BC V1V 1V7, CANADA

**Abstract**

Sol argues that innovation propensity is not a specialized adaptation resulting from targeted selection but an instance of exaptation because selection cannot act on situations that are only encountered once. In exaptation, a trait that originally evolved to solve one problem is co-opted to solve a new problem; thus the trait or traits in question must be necessary and sufficient to solve the new problem. Sol claims that traits such as persistence and neophilia, are necessary and sufficient for animal innovation, which is a matter of trial and error. We suggest that this explanation does not extend to human innovation, which involves strategy, logic, intuition, and insight, and requires traits that evolved, not as a byproduct of some other function, but for the purpose of coming up with adaptive responses to environmental variability itself. We point to an agent based model that indicates the feasibility of two such proposed traits: (1) *chaining,* the ability to construct complex thoughts from simple ones, and (2) *contextual focus,* the ability to shift between convergent and divergent modes of thought. We agree that there is a sense in which innovation is exaptation—it occurs when an existing object or behaviour is adapted to new needs or tastes—and refer to a mathematical model of biological and cultural exaltation. We conclude that much is gained by comparing and contrasting animal and human innovation.

The aim of Sol's provocative chapter is to develop a comprehensive framework for the evolution of animal innovation. The framework can be summarized as follows:

- Innovation propensity is a by-product of a combination of traits including motivation, emotional responses, cognitive abilities, and morphological constraints.
- These traits initially evolved for other functions and were co-opted for innovative problem solving through exaptation.
- Through genetic assimilation, learned traits—such as those that underlie innovation propensity—may eventually become innate.

In this commentary we discuss this framework and (as instructed by the editors) take the ball and run with it.

## Innovative Capacity as Exaptation

The rationale for Sol's argument (that innovation propensity is not a specialized adaptation resulting from targeted selection but an instance of exaptation, i.e., a by-product of selection for other traits) is that selection cannot act on situations that are only encountered once. (You only have to innovate once because if you encounter that situation again you can simply *remember* what you did the first time.) He proposes that all natural selection can do to prepare you to seize the moment and act on affordances for innovation when they present themselves is provide you with the following general characteristics: (1) *motivation,* e.g., hunger may increase persistence in finding ways to obtain food, (2) *emotional responses,* e.g., neophilia versus neophobia, (3) *cognitive abilities* such as attention, discrimination, and the capacity for episodic learning, (4) *morphological and physiological constraints* on the type and diversity of motor patterns, including the ability to adjust actions in response to context and (5) *time*, e.g., a longer lifespan provides more time for problem solving. It is these traits, Sol argues, with their interacting constraints and trade-offs (which are in some cases functionally linked due to pleiotropic genetic effects and common physiological pathways) that have been the target of selection over the life history of an organism, not innovativeness itself.

We agree that these traits contribute to the capacity to innovate. We also believe they contribute to the capacity to find food, find a mate, care for offspring, and so forth, yet no one would argue that mate-finding or offspring care are exaptations. In exaptation (sometimes called preadaptation), a trait that originally evolved to solve one problem is co-opted to solve a new problem; the trait or traits in question must be necessary *and sufficient* to solve the new problem. Are persistence, neophilia, and so forth sufficient for innovation? Sol argues that this is so for innovation in animals, which he claims is generally a matter of trial and error. As researchers who focus on human innovation, we point out that while persistence, neophilia, and so forth, are necessary they are not sufficient for the strategic, logical, intuitive, insightful, and even therapeutic processes involved in considering and reconsidering a complex idea from different real and imagined perspectives until all the bits and pieces fall into place.

Let us reconsider the starting point for Sol's argument: that selection cannot act on situations that are only encountered once. Viewed at a sufficiently fine level of granularity *all* situations are new situations that have never been encountered before (as Heraclitus said, you never step into the same river twice). Conversely, viewed at a sufficiently coarse level of granularity, all situations have been encountered previously. The issue then is: do novelty-affording situations collectively have enough in common at some intermediate level of granularity for selection to act upon? For example, is there a trait (or traits) that evolved, not as a byproduct of some other function, but expressly for the purpose of coming up with innovative, adaptive responses to environmental variability itself?

We believe the answer is yes. This position is supported by experiments carried out using a computational model of cultural evolution that showed that the mean fitness of ideas across a society of artificial agents increases with the introduction of two innovation enhancing abilities: (1) *chaining*, the ability to combine simple ideas into complex ones, and (2) *contextual focus*, the ability to shift from a convergent to a divergent processing mode when the fitness of one's current actions is low (Gabora & Saberi, 2011; Gabora & DiPaola, 2012). Moreover, both factors—chaining and contextual focus—proved most useful in times of environmental fluctuation (Gabora, Chia, & Firouzi, 2013). Of course, care must be taken in extrapolating from a simple computational model to the real world. However, the computer experiments are not the only source of support; Chrusch and Gabora (2014) synthesized these computational modeling results with findings from behavioural genetics, psychology, and anthropology to produce an integrated multi-level account of how chaining, contextual focus, and thereby human creative abilities could have evolved.

In short, these additions enable Sol's basic argument to be extended from trial and error innovative problem solving in animals to the more complex innovative abilities exhibited by humans.

## The Role of Context

Sol stresses that the extent to which an animal expresses its capacity to innovate may depend on the relative costs and benefits of different actions *in its particular ecological context*; for example, populations of a given species living in more variable environments tend to be more innovative than those in predictable environments. Thus, he says, innovativeness hinges on not just the environment per se, but on the animal's *interaction* with its environment. We agree, and suggest that it has in this sense that innovation is exaptation. Gabora, Scott, and Kauffman (2013) developed a mathematical framework for exaptation with examples from both biological evolution and the evolution of cultural novelty through innovation. It is actually a quantum model, not in the sense of Penrose, but in the sense that it uses a generalization of the quantum formalism that was developed to model situations involving extreme contextuality in the macroworld. The state of a trait (or the starting point for an idea) is written as a linear superposition of a set of basis states, or possible forms the trait (or idea) could evolve into, in a complex Hilbert space. (For example, the basis states might represent possible ways of using a tire.) These basis states are represented by mutually orthogonal unit vectors, each weighted by an amplitude term. The choice of possible forms (basis states) depends on the context-specific goal or adaptive function of interest, which plays the role of an observable. (For example, in the context of wanting to create a playground someone turned a useless tire into a tire swing.) Observables are represented by self-adjoint operators on the Hilbert space. The possible forms (basis states) corresponding to this adaptive function (observable) are called eigenstates. In this model, innovative capacity did not evolve as an exaptation from some other, selected-for adaptive trait. Rather, innovation itself—or at least the retooling of an object or idea by considering it from a new point of view—is modeled as exaptation.

## The Role of Genetic Assimilation

As Sol points out, when innovations are essential for survival, the nexus of traits underlying innovative capacity become canalized. A phenotypic response to an environmental condition, such as a learned innovative behavior, can over time be *genetically assimilated*, and thus innate.

Some limitations of innate behavior are (1) it is rather unflexible, and (2) it operates over the course of biological generations. Thus while some kinds of innovation may be genetically assimilated it is unlikely that the innovations that fuel human cultural evolution are, given that they can unfold spontaneously over timeframes of hours or minutes (e.g., humorous internet banter).

## Innovation as Viewed by the Animal Behavior Literature versus the Psychological Literature

We found it interesting to compare and contrast how innovation is viewed from the animal behavior literature versus the psychological literature. Sol's four stages in the innovation process—sampling, exploring, problem solving, and learning (by which mean means incorporating the solution into a behavioural repertoire)—bear some resemblance to Wallas' (1926) four stages of the creative process: preparation, incubation, illumination, and verification. The notion of "sampling" appears to be related to the notion of "problem finding" (Getzels & Csikszentmihalyi, 1976; Mumford, Reiter-Palmon, & Redmond, 1994; Runco & Chand, 1994), and the first two stages map onto the "generate" and explore" stages of the creative cognition approach (Ward, 1995). What Sol refers to as *neophilia* seems comparable to the human personality trait of 'openness to experience'. Sol notes that innovativeness may be related to risk taking, and there is indeed evidence that highly creative individuals tend to take more risks (Merrifield et al., 1961).

We note, however, that there are also differences in how innovativeness is viewed by these two fields. While Sol's focus is squarely on innovative problem solving (e.g., opening a lid to find hidden food), psychologists who seek a general scientific framework for creativity often unite innovative problem solving under the same broad umbrella as abilities such as art-making and scientific theorizing. Through Sol's chapter we came to better appreciate how by comparing and contrasting simple versus complex forms of innovation we sharpen our understanding of how new objects and forms of behavior come to be.

## Acknowledgements

This research was supported in part by a grant from the Natural Sciences and Engineering Research Council of Canada.